\documentclass{article}
\usepackage{graphicx} 
\usepackage{algorithm}
\usepackage{algpseudocode}

\title{ChargingBoul: A Competitive Negotiating Agent with Novel Opponent Modeling}
\author{Joe Shymanski}

\begin{document}

\maketitle

\begin{abstract}
    Automated negotiation has emerged as a critical area of research in multiagent systems, with applications spanning e-commerce, resource allocation, and autonomous decision-making. This paper presents ChargingBoul, a negotiating agent that competed in the 2022 Automated Negotiating Agents Competition (ANAC) and placed second in individual utility by an exceptionally narrow margin. ChargingBoul employs a lightweight yet effective strategy that balances concession and opponent modeling to achieve high negotiation outcomes. The agent classifies opponents based on bid patterns, dynamically adjusts its bidding strategy, and applies a concession policy in later negotiation stages to maximize utility while fostering agreements. We evaluate ChargingBoul’s performance using competition results and subsequent studies that have utilized the agent in negotiation research. Our analysis highlights ChargingBoul’s effectiveness across diverse opponent strategies and its contributions to advancing automated negotiation techniques. We also discuss potential enhancements, including more sophisticated opponent modeling and adaptive bidding heuristics, to improve its performance further.
\end{abstract}

\section{Introduction}

Automated negotiation has long been a critical area of research in artificial intelligence, with applications ranging from e-commerce and diplomatic mediation to resource allocation in multiagent systems. The Automated Negotiating Agents Competition (ANAC) serves as a benchmark for advancing the field, challenging participants to develop autonomous agents capable of strategic, adaptive, and efficient negotiation.

In 2022, we entered ChargingBoul into the Automated Negotiation League (ANL) of ANAC, where it competed against dozens of agents submitted by teams from around the world. The competition featured bilateral multiattribute negotiations, where each agent sought to maximize its utility while interacting with diverse opponents. Each negotiation session consisted of 50 sequential, timed rounds, with agents alternately exchanging bids based on predefined preference profiles. The final agreement, if reached before time expired, determined the agents' respective utilities, calculated via a linear additive utility function.

Performance in the competition was assessed using two primary metrics:
\begin{enumerate}
    \item \textbf{Individual Utility}: The average utility an agent secured across negotiations.
    \item \textbf{Social Welfare}: The sum of utilities between both negotiating agents per round.
\end{enumerate}

ChargingBoul proved to be highly competitive, securing second place in individual utility, falling short of first place by only 0.001 utility points on average. Its success stemmed from a straightforward yet effective negotiation strategy, leveraging opponent modeling and adaptive bidding techniques. This paper details the design choices that contributed to ChargingBoul’s performance, compares its results to previous approaches, and discusses potential improvements for future autonomous negotiating agents.

\section{Literature Review}

Automated negotiation has been widely studied within the domains of multiagent systems and game theory. Early approaches to autonomous negotiation were inspired by classical economic models, such as Nash bargaining solutions and concession-based tactics. More recent research has emphasized adaptive strategies that leverage opponent modeling, machine learning, and utility-driven decision-making.

The ANAC competition has provided a valuable testing ground for these advancements, with previous studies analyzing both individual agent strategies and broader trends in competitive negotiation. Notable approaches include de Jonge's MiCRO strategy~\cite{de2022analysis}, the ANOTO agent~\cite{chen2024anoto}, and Luckyagent2022~\cite{ebrahimnezhad2022luckyagent2022}, which have demonstrated the effectiveness of heuristic-based bidding, learning-based adaptation, and opponent classification techniques.

Like many of its predecessors, ChargingBoul builds upon existing methodologies, employing a structured opponent modeling framework alongside a dynamically adaptive strategy. Additionally, ChargingBoul introduces two novel statistics by which it classifies opponent negotiation tactics and subsequently adjusts its bidding strategy. By evaluating its performance within ANAC 2022, this paper contributes to the ongoing discourse on effective autonomous negotiation and explores how lightweight yet effective strategies can achieve competitive results.

\section{Methodology}

\begin{figure}
    \centering
    \includegraphics[width=0.75\linewidth]{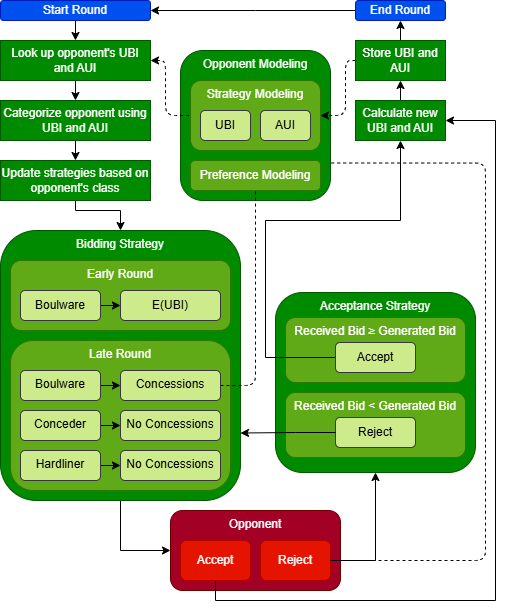}
    \caption{Overview of ChargingBoul's entire negotiation strategy.}
    \label{fig:entire-strategy}
\end{figure}

An automated negotiation agent typically contains three main components: its own bidding strategy, an acceptance strategy, and a method for modeling the opponent's bidding strategy. ChargingBoul implements all three. We will present its novel opponent modeling first to introduce a few key terms that are necessary to understand the other two mechanisms of the agent. Figure~\ref{fig:entire-strategy} presents an overview of the whole strategy.

\subsection{Opponent Modeling}

There are two ways in which to model the opponent's behavior: preference modeling and strategy modeling. The former approximates the opponent's preferred issues and values given the current negotiation domain. This allows the agent to estimate the utility they receive from any proposed bid. The latter estimates the opponent's bidding strategy throughout a single round of negotiation.

\subsubsection{Preference Modeling}

In order to predict the opponent's utility from a given bid, we employed a frequency-based scheme for preference modeling. The model was updated within the round of negotiation with each received bid. Being able to predict the opponent's utility is useful for determining when to make concessions later in the round, which we will discuss later.

\subsubsection{Strategy Modeling}

We created two novel statistics for estimating the opponent's bidding strategy: unique bid index (UBI) and average utility index (AUI). Both values are computed and saved at the end of the round using the sequence of bids proposed by the opponent. They are then retrieved at the start of the next round to update our agent's bidding strategy.

\paragraph{UBI} This value is used to estimate when the opponent begins to offer new bids. It is calculated using a recursive process. The agent looks at the sequence of bids it received from the opponent and computes the number of unique bids in the left and right halves. If this number is larger for the right half, the index is incremented and the process repeats on the right half of bids. The pseudocode is given by Algorithm~\ref{alg:ubi}, where the initial call passes in the list of received bids as $receivedBids$ and sets $ubi=0$.

\begin{algorithm}
\caption{Calculating unique bid index (UBI)}
\label{alg:ubi}
\begin{algorithmic}
    \Function{calculateUBI}{$receivedBids$, $ubi$}
        \State $left \gets$ left half of $receivedBids$
        \State $right \gets$ right half of $receivedBids$
        \State $lenLeft \gets$ number of unique bids in $left$
        \State $lenRight \gets$ number of unique bids in $right$
        \If{$lenLeft > 0$ and $lenRight > 0$ and $lenLeft < lenRight$}
            \State $ubi \gets$ \Call{calculateUBI}{$right$, $ubi+1$}
        \EndIf
        \State \Return $ubi$
    \EndFunction
\end{algorithmic}
\end{algorithm}

\paragraph{AUI} The next metric is critical for estimating how much the opponent is conceding as the round progresses. This is evaluated using a similar recursive process. The mean utilities from the first and last halves of the opponent's bids are compared. If the right half contains a higher average utility, the AUI is incremented and the process continues with the right interval. Algorithm~\ref{alg:aui} contains this procedure's pseudocode. Initially, $receivedUtils$ corresponds to the list of bid utilities received from the opponent's offers, and $aui$ is set to 0.

\begin{algorithm}
\caption{Calculating average utility index (AUI)}
\label{alg:aui}
\begin{algorithmic}
    \Function{calculateAUI}{$receivedUtils$, $aui$}
        \State $left \gets$ left half of $receivedUtils$
        \State $right \gets$ right half of $receivedUtils$
        \State $lenLeft \gets$ number of utilities in $left$
        \State $lenRight \gets$ number of utilities in $right$
        \State $meanLeft \gets$ mean of $left$
        \State $meanRight \gets$ mean of $right$
        \If{$lenLeft > 0$ and $lenRight > 0$ and $meanLeft < meanRight$}
            \State $aui \gets$ \Call{calculateAUI}{$right$, $aui+1$}
        \EndIf
        \State \Return $aui$
    \EndFunction
\end{algorithmic}
\end{algorithm}

ChargingBoul looks up both of these statistics at the start of each negotiation. It uses them to characterize the opponent's strategy into one of three categories: Boulware, Hardliner, and Conceder. If $ubi \geq 5$, then the opponent's strategy is classified as Boulwarish. If not, the opponent is a Hardliner when $aui \leq 2$ and a Conceder otherwise. Our agent then changes its bidding strategy according to the opponent's category.

\subsection{Bidding Strategy}

Our agent employs a default strategy during its first negotiation with an opponent. It then updates its bidding strategy in the current round based on the opponent's behavior from the previous round.

\subsubsection{Early Round}

The ChargingBoul's bidding strategy is Boulwarish on average, with an interval from which the proposed bid is randomly chosen. As time increases, the utility goal decreases while this interval of bids widens. We choose bids randomly to explore the bid space while making our bidding strategy harder to detect.

The utility goal $g(t)$ is given by Equation~\ref{eq:util_goal}.
\begin{equation}
    \label{eq:util_goal}
    g(t) = m+(1-m)(1-t^{1/E})
\end{equation}
The minimum acceptable utility ($m$) denotes the lowest value that the utility goal will reach at the end of the round when $t=1$, while $E$ controls the shape of the concession curve. Smaller $E$ values delay concessions further into the round. The default values for $m$ and $E$ are .5 and .1, respectively. When facing a Conceder, $m$ is set to .4 to further ensure that a deal is made.

\begin{figure}
    \centering
    \includegraphics[width=0.75\linewidth]{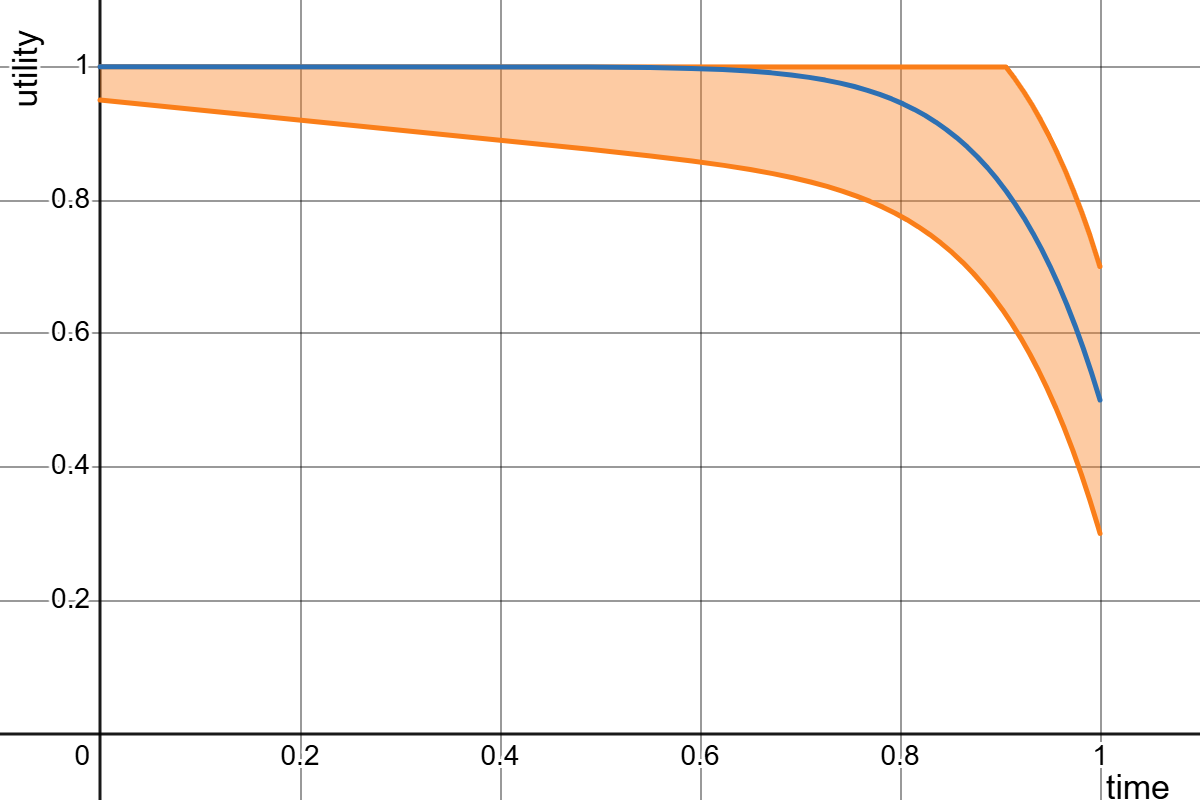}
    \caption{Default bidding strategy ($m=.5$, $E=.1$, $\epsilon=.05$).}
    \label{fig:bidding-strategy}
\end{figure}

Equation~\ref{eq:util_bounds} defines the range from which a random bid is selected.
\begin{equation}
    \label{eq:util_bounds}
    [g(t)-(3t+1)\epsilon,~g(t)+(3t+1)\epsilon]
\end{equation}
The ``tolerance'' variable ($\epsilon$) captures the granularity of the bid space. A domain with fewer bids, and thus a larger $\epsilon$, requires a wider interval to encapsulate an acceptable number of bids. For our tests, $\epsilon$ typically falls between .001 and .05. Figure~\ref{fig:bidding-strategy} demonstrates the agent's default bidding strategy during the course of a single negotiation.

If ChargingBoul selects a bid with lower utility than one it has already received and rejected, it will re-propose the rejected bid with the highest utility. In this way, the agent never undersells itself, and the opponent should accept any bids it has previously proposed in the negotiation.

Some of the parameters change slightly at the beginning of each round, depending on the classification of the opponent's strategy from the previous negotiation. ChargingBoul's concession severity adapts to Boulwarish opponents according to Equation~\ref{eq:e}.
\begin{equation}
    \label{eq:e}
    E = .2\cdot2^{5-ubi}
\end{equation}
If the UBI from the previous round is large, we estimate that the opponent conceded very late in the round. Thus, we change $E$ for the next negotiation to concede just as late, so we are not exploited for conceding too early. This ensures that our concession curve looks similar to the opponent's, so neither agent has an unfair advantage.

\subsubsection{Late Round}

\begin{figure}
    \centering
    \includegraphics[width=0.75\linewidth]{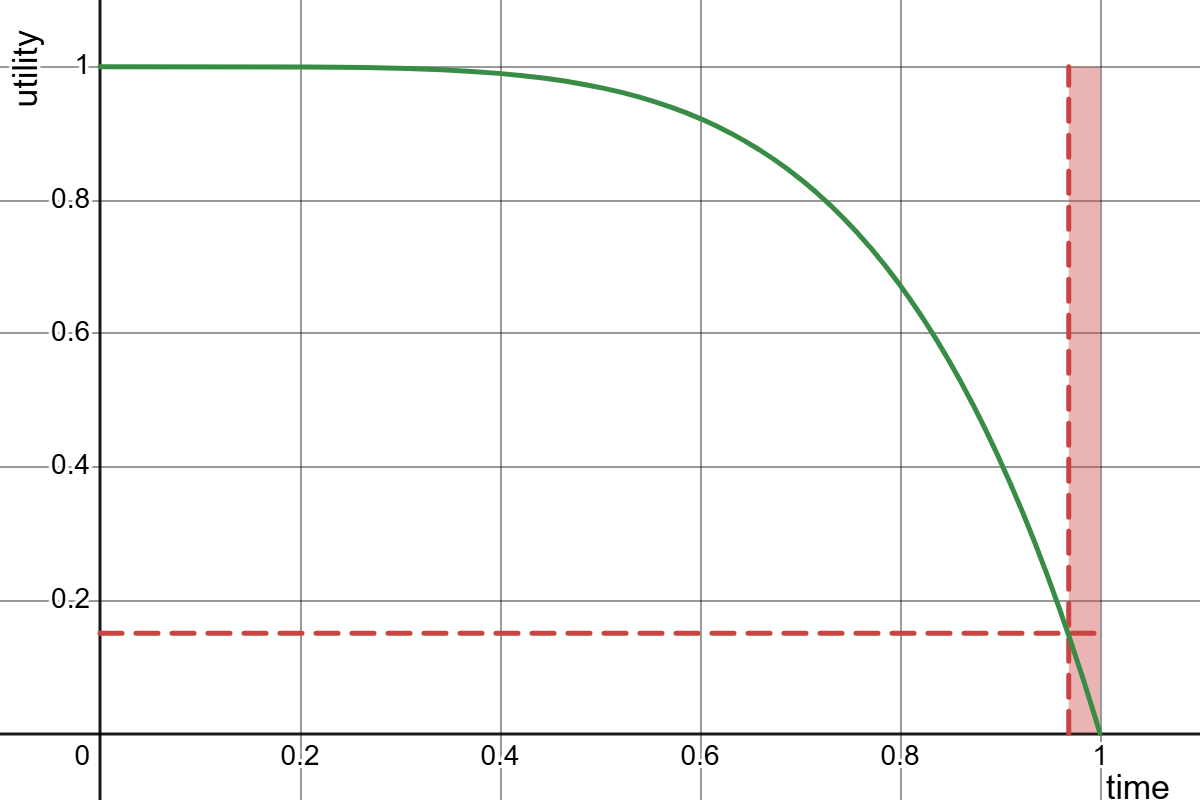}
    \caption{Estimated opponent strategy ($ubi=5$) and 15\% concession period.}
    \label{fig:opponent-strategy}
\end{figure}

If the round is nearing its conclusion and an agreement has yet to be made, ChargingBoul will make drastic concessions based on the opponent's strategy. This period is defined in Equation~\ref{eq:end}.
\begin{equation}
    \label{eq:end}
    t > 1-.5^{ubi}
\end{equation}
This represents the point in the negotiation when we believe the opponent is within $\sim$15\% of their minimum acceptable utility, demonstrated by Figure~\ref{fig:opponent-strategy}. For Boulwarish opponents, our agent first lowers $m$ from .5 to .3. If the best received utility so far is greater than $m$ and the predicted opponent utility is less than $2m$, then ChardingBoul re-proposes the bid. Otherwise, a new bid is generated like normal. This concession strategy protects our agent from being overly exploited by stubborn negotiators throughout the tournament and instead punishes those who do.

Currently, the agent does not make concessions for other opponent strategies. This is because Hardliners only succeed when they overly exploit concession-making opponents, so reaching no agreement hurts them more than it hurts ChargingBoul in a tournament. Conceders, on the other hand, should eventually come to an agreement without an increase in concession-making.

\subsection{Acceptance Strategy}

Our agent is willing to accept an appropriate offer at any time throughout the negotiation. However, CharingBoul always generates a bid upon receiving an offer in order to compare the two. If the received bid is better than the one it generated, the opponent's offer is accepted, concluding the round. Otherwise, the offer is rejected and the new bid is proposed.

\section{Results}

In the 2022 ANAC ANL competition, CharingBoul finished runner-up in average utility (.724)~\cite{aydougan202213th}. This was its first official contest, where it faced dozens of negotiating agents from around the world, and it finished only .001 utility points behind the winner.

One of the negotiators submitted to that tournament (LuckAgent2022) had a bug in its code, preventing it from reaching agreements in most negotiations~\cite{aydougan202213th}. The creators of this agent fixed the error, and the organizers ran another (unofficial) iteration of the tournament after its conclusion~\cite{ebrahimnezhad2022luckyagent2022}. ChargingBoul obtained the highest average utility (.742) in this tournament. Even when the agents were stripped of their learning mechanisms, our agent still ranked first.

The creators of ANOTO, an agent that utilizes offline-to-online reinforcement learning, tested their automated negotiator against several notable performers in the literature and past ANL tournaments~\cite{chen2024anoto}. ChargingBoul finished tenth in average utility (.62).

Renting \textit{et al.} created a GNN- and RL-based negotiating agent, which performed well against a handful of simple baseline models~\cite{renting2024towards}. However, when pitted against many of the negotiators from the 2022 ANL competition, it seemed to falter. ChargingBoul obtained the highest average utility ($\sim$.96) against their agent out of all the competitors tested.

\section{Conclusion}

In this paper, we presented ChargingBoul, a negotiating agent designed for the Automated Negotiation League (ANL) of ANAC 2022. Despite its relatively simple approach, ChargingBoul demonstrated remarkable effectiveness, securing second place in individual utility, narrowly missing the top spot by only 0.001 utility points. Its success highlights the power of structured opponent modeling and adaptive bidding in competitive multiattribute negotiations.

ChargingBoul’s strategy balances competitiveness and cooperation, optimizing its utility while engaging with diverse opponents across hundreds of negotiation rounds. The insights gained from its performance contribute to the broader understanding of agent design in automated negotiation, particularly in environments with limited prior knowledge of opponents.

While ChargingBoul achieved strong results, there remain avenues for further refinement and experimentation. In the next section, we outline several promising directions for enhancing automated negotiating agents beyond the strategies employed here.

\section{Future Work}

Although ChargingBoul was highly competitive, there are several areas where its methodology could be extended:
\begin{enumerate}
    \item \textbf{Better Class-Specific Responses}: The agent can change its bidding strategy more drastically for Conceder and Hardliner opponents.
    \item \textbf{Enhanced Opponent Modeling}: ChargingBoul currently classifies opponents into three broad behavioral categories. More granular modeling techniques, such as reinforcement learning-based adaptation or Bayesian opponent classification, could further refine its strategic responses.
    \item \textbf{Dynamic Strategy Adjustment}: While ChargingBoul adapts over multiple rounds, future iterations could incorporate real-time learning to dynamically adjust bidding strategies within a single negotiation session. This could involve detecting shifts in an opponent’s behavior and adjusting accordingly.
    \item \textbf{Exploring Alternative Utility Functions}: The competition used a linear additive utility function, but real-world negotiation scenarios often involve nonlinear or multi-objective utility landscapes. Extending ChargingBoul to handle such cases would increase its applicability.
    \item \textbf{Multiagent Negotiation and Coalition Formation}: While ANL focuses on bilateral negotiations, many real-world applications require agents to negotiate with multiple parties simultaneously or form temporary coalitions. Extending ChargingBoul’s strategy to multiagent settings could open up new research directions.
    \item \textbf{Long-Term Learning and Meta-Strategy Optimization}: Incorporating a meta-learning framework that allows ChargingBoul to retain and refine knowledge across multiple tournaments could improve its ability to generalize against unseen opponents.
    \item \textbf{Application to Real-World Domains}: While designed for ANAC, ChargingBoul’s negotiation techniques could be adapted for practical applications such as automated contract negotiation, e-commerce bargaining, and resource allocation in multiagent systems. Future research could explore how well its strategy translates beyond controlled competition settings.
\end{enumerate}
By pursuing these enhancements, we aim to push the boundaries of automated negotiation research and further bridge the gap between theoretical models and practical deployment.

\section{Acknowledgments}

We would like to thank Diego Velasco and Selim Karaoglu for their contributions to the development of ChargingBoul.

\bibliographystyle{plain}
\bibliography{references}

\end{document}